\documentclass[DIV=calc, paper=a4, fontsize=11pt, twocolumn]{scrartcl}	 

\usepackage[english]{babel} 
\usepackage[protrusion=true,expansion=true]{microtype} 
\usepackage[svgnames]{xcolor} 
\usepackage[hang, small,labelfont=bf,up,textfont=it,up]{caption} 
\usepackage{booktabs} 
\usepackage{fix-cm}	 

\usepackage{hyperref}
\usepackage[pdftex]{graphicx}

\usepackage{sectsty} 
\allsectionsfont{\usefont{OT1}{phv}{b}{n}} 

\usepackage{fancyhdr} 
\usepackage{lastpage} 

\usepackage{lettrine} 
\newcommand{\initial}[1]{ 
\lettrine[lines=3,lhang=0.3,nindent=0em]{
\color{DarkGoldenrod}
{\textsf{#1}}}{}}

\usepackage{titling} 

\newcommand{\HorRule}{\color{DarkGoldenrod} \rule{\linewidth}{1pt}} 

\pretitle{\vspace{-30pt} \begin{flushleft} \HorRule \fontsize{50}{50} \usefont{OT1}{phv}{b}{n} \color{DarkRed} \selectfont} 

\title{Impact assessment for vulnerabilities in open-source software libraries}

\posttitle{\par\end{flushleft}\vskip 0.5em} 

\preauthor{\begin{flushleft}\large \lineskip 0.5em \usefont{OT1}{phv}{b}{sl} \color{DarkRed}} 

\author{Henrik Plate, Serena Elisa Ponta, Antonino Sabetta, } 

\postauthor{\footnotesize \usefont{OT1}{phv}{m}{sl} \color{Black} 
SAP Labs France 

\par\end{flushleft}\HorRule} 

\date{10 April 2015} 

\begin{document}

\maketitle 

\thispagestyle{fancy} 


\initial{S}\textbf{oftware applications integrate more and more open-source software
(OSS) to benefit from code reuse. As a drawback, each vulnerability
discovered in bundled OSS potentially affects the application. Upon
the disclosure of every new vulnerability, the application vendor has
to decide whether it is exploitable in his particular usage context,
hence, whether users require an urgent application patch containing a
non-vulnerable version of the OSS. Current decision making is mostly
based on high-level vulnerability descriptions and expert knowledge,
thus, effort intense and error prone. This paper proposes a pragmatic
approach to facilitate the impact assessment, describes a
proof-of-concept for Java, and examines one example vulnerability as
case study. The approach is independent from specific kinds of
vulnerabilities or programming languages and can deliver immediate
results.}


\section{Introduction}
\label{sec:intro}
The adoption of open-source software (OSS) in the software industry
has continued to grow over the past few years and many of today's
commercial products are shipped with a number of OSS libraries.
Vulnerabilities of any of these OSS libraries can have considerable
consequences on the security of the commercial product that bundles
them. 
The relevance of this problem has been acknowledged by \textsf{OWASP} which
included ``A9-Using Components with Known Vulnerabilities'' 
among the Top-10 security vulnerabilities in 2013~\cite{owasptop10}.
The disclosure in 2014 of vulnerabilities such as
Heartbleed\footnote{\url{http://heartbleed.com/}}
and ShellShock\footnote{\url{https://shellshocker.net/}}
contributed even further to raise the awareness of the problem.

Despite the deceiving simplicity of the existing solutions 
(the most obvious being: update to a more recent, patched version),
OSS libraries with known vulnerabilities are found to be used for some
time after a fixed version has been issued~\cite{contrast-security}.  
Updating to a more recent, non-vulnerable version of a library
represents a straightforward solution at development time. However, the problem
can be considerably more difficult to handle when a vulnerable OSS
library is part of a system that has already been deployed and made
available to its users. In the case of large enterprise systems that
serve business-critical functions, any change (including updates) may
cause system downtime and comes with the risk that new unforeseen
issues arise.
For this reason, software vendors need to carefully assess whether an application
requires an urgent application patch to update an OSS library, or whether the
update can be done as part of the application's regular release cycle.

The key question that vendors have to answer is whether or not a given
vulnerability, that was found in an OSS library used in one of their
products, is indeed exploitable given the particular use that such a product
makes of that library~\cite{younis2014using}. If the answer is
positive, an application patch must be produced, and its installation needs to be triggered for all existing deployments of the application. 
If the answer is negative, the library update can be scheduled as part of the
regular release cycle, without causing extra efforts related to
urgent patch production by vendors and patch installation by users.
The current practice of assessing the potential
impact of a vulnerability in OSS is time-consuming and
error-prone. Vulnerabilities are typically documented
with short, high-level textual descriptions expressed in natural
language; at the same time, the assessment demands considerable expert
knowledge about the application-specific use of the library in
question.
Consequences of wrong
assessments can be expensive: if the developer wrongly assesses that a
given vulnerability is not exploitable, application users remain
exposed to attackers.
If she wrongly judges that it is exploitable, the effort of
developing, testing, shipping, and deploying the patch to the customers' systems is spent in vain.

This paper presents a pragmatic approach that contributes to simplify
the decision making.
We do so by automatically producing (whenever possible) concrete
evidences supporting the case for urgent patching. More specifically
we assess whether an application uses (portion of) a library for which a
security fix has been issued in response to a vulnerability.
The approach seamlessy integrates in the usual development
workflow without requiring additional effort from developers and is
independent of programming languages and vulnerability types.

The paper is organized as follows. Section~\ref{sec:concept}
presents our approach and a generic architecture that supports it.
Section~\ref{sec:eval} describes our prototypical implemention and its
application to a case study. Section~\ref{sec:quality} outlines 
observations regarding the integration and quality of information stemming from different sources.
Section~\ref{sec:relwork} presents related literature. Section~\ref{sec:concl} concludes the
paper.

\section{Approach}
\label{sec:concept}

\subsection*{Concept}
\label{subsec:concept}

\begin{figure*}[t!]
\centering
\includegraphics[width=.9\textwidth]{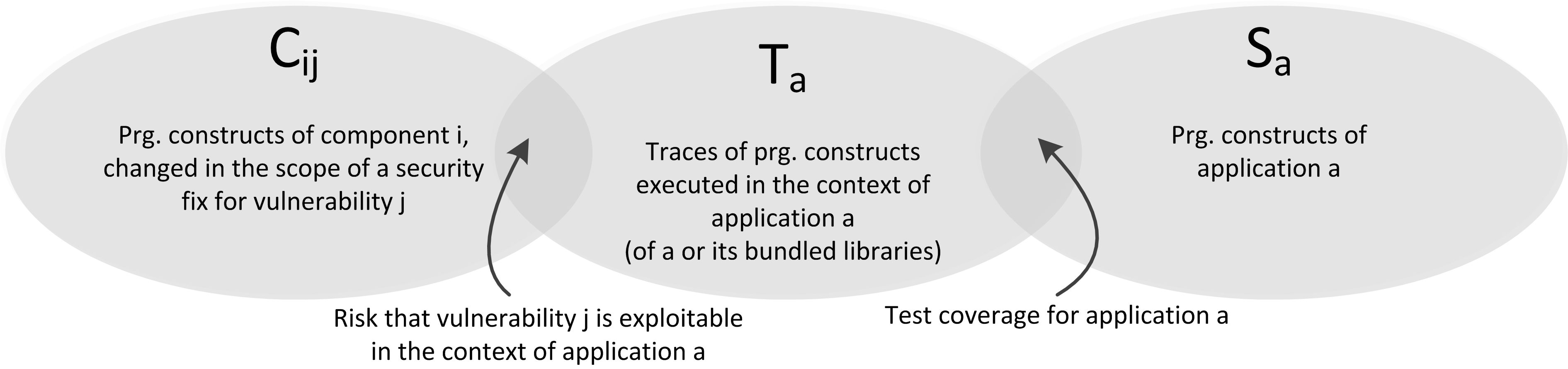}
\caption{Main concept}
\label{fig:concept}
\end{figure*}

In order to assess whether or not a given vulnerability in an OSS
library is relevant for a particular application, we
consider the corresponding \emph{security patch}, i.e., the set of changes
performed in the source code of the library in response
to the vulnerability. Our approach is than based on the 
following pragmatic assumption:
\begin{itemize}
\item[{\bf (A1)}] Whenever an application that includes a library (known
to be vulnerable) executes a fragment of the library that would be updated
in a security patch, there exists a significant risk that the vulnerability
can be exploited.
\end{itemize}
The underlying idea is that if programming constructs\footnote{We use the language-agnostic term ``programming construct'' to refer to structural elements such as methods, constructors, functions and so on.} that
would be changed by the patch are used, than the application is using code involved in the
vulnerable part of the library. Therefore, we collect execution
traces of applications, and compare those with changes that would be introduced
by the security patches of known vulnerabilities in order to detect
whether ``critical'' library code is executed. Figure~\ref{fig:concept}
illustrates the approach graphically: The change-list $C_{ij}$
represents the set of all programming constructs of OSS component $i$ that were modified, added or
deleted as part of the security patch for vulnerability $j$.
Change-lists can be computed as soon as a security patch has been
produced for a vulnerability. Patches can be assumed available at the
time vulnerabilities become public in the case of responsible
disclosure. The trace-list $T_{a}$ represents the set of all
programming constructs, either part of application $a$ or any of its bundled
libraries, that were executed at least once during the runtime of
application $a$.  The collection of traces can be done at many
different times, starting from unit tests 
until the application is deployed for productive use.


The intersection $C_{ij} \cap T_{a}$ comprises all those programming constructs
that are both subject to security patch $j$ and have been executed in the
context of application $a$.
Following assumption {\bf (A1)}, a non-empty intersection $C_{ij}\cap T_{a}$
indicates that a newly disclosed vulnerability is highly relevant, due to the
risk of exploitability. An empty intersection, on the other hand, may result
from insufficient coverage, hence, it does not automatically render a
vulnerability irrelevant. Coverage is described by the intersection
of the sets $T_{a}\cap S_{a}$, where  $S_{a}$ is the set of all programming
constructs belonging to the application itself. The larger the intersection, the
better the coverage, and the greater the confidence that constructs belonging to
$C_{ij}$ cannot be reached.

Library versions can be disregarded, provided that the traces were
collected before the release of a security patch 
and that all existing library versions are affected by the vulnerability. In other
cases (in particular if traces are more recent than a patch, and if the
corresponding constructs exist in both vulnerable and patched library version)
one has to identify and compare the version of the library producing the trace with the
ones affected by the vulnerability. 


Our approach is not immune to reporting false-positives and
false-negatives.  False-positives occur if a vulnerability is not
exploitable despite a non-empty intersection $C_{ij} \cap T_{a}$,
which is due to the fact that exploitability can depend on many other
conditions, e.g., the presence of sanitization techniques in the
application or specific configuration settings. 
False-negatives occur if vulnerabilities are exploitable despite an
empty intersection $C_{ij}\cap T_{a}$, which can result from
insufficient coverage, a problem shared with many techniques relying on
dynamic execution (as opposed to static analysis).
The silver-lining is that the approach is entirely independent of programming
languages or types of vulnerabilities and it can provide immediate
results. Intuitively, we expect that developers are convinced to
update a library when presented with a non-empty intersection $C_{ij}
\cap T_{a}$, i.e., traces of programming constructs that are subject
to a security patch. In any case, the information collected
is valuable to simplify further analysis, complementing the
high-level vulnerability description expressed in natural language in
publicly available vulnerability databases, such as the National Vulnerability
Database.

\subsection*{Architecture.} \label{sec:arch}
\begin{figure*}[t]
\centering
\includegraphics[width=.85\textwidth]{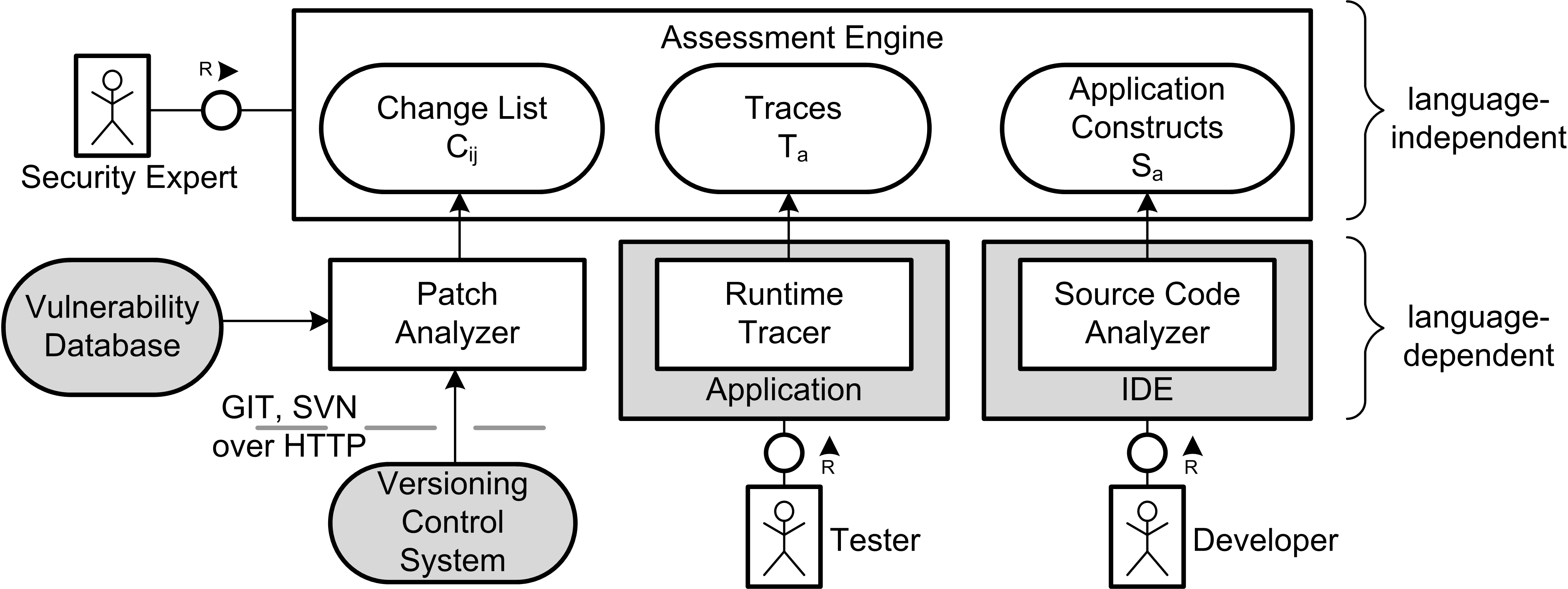}
\caption{Generic Solution Architecture}
\label{fig:arch}
\end{figure*}

Figure~\ref{fig:arch} illustrates a generic architecture that supports
our approach.
Components depicted in white belong to the proposed solution and require an implementation, while components depicted in
grey represent the solution's environment. A specific implemention for
Java and related tooling is described in Section~\ref{sec:eval}.

The Assessment Engine on top is responsible for storing and aggregating the
three sets of Figure~\ref{fig:concept}. 
It also presents assessment results concerning the relevance of
vulnerabilities to the security expert of the respective
application. 

The Patch Analyzer is triggered upon the publication of a new
vulnerability for an open-source library. It interacts with the
respective Versioning Control System (VCS), identifies all programming
constructs changed to fix the vulnerability, and uploads their
signatures to the central Assessment Engine.  This change-list is
built by comparing the vulnerable and patched revision of all relevant
source code files of the library. The corresponding commit revisions
can be obtained from the vulnerability database, searched in the
commit log, or specified manually.

The Runtime Tracer collects execution traces of programming
constructs, and uploads them to the central engine. This is achieved
by injecting instrumentation code into all programming constructs of
the application itself and all the bundled libraries. 
The instrumentation can be done dynamically,
during the actual runtime, or statically, prior to the application's
deployment. The former can guarantee the tracing of all programming
constructs used at runtime including, e.g., libraries included at runtime and 
parts of the runtime environment. Its major drawback is the impact on
application startup, in particular if many contructs need to be loaded
before the applicaton becomes available to its users, as in the case
of application containers. Static instrumentation
does not impact the application startup time and can be used in cases
where the runtime environment cannot be configured for dynamic
instrumentation, e.g., in PaaS environments. However, it cannot
guarantee the coverage of programming constructs used at runtime.

The Source Code Analyzer scans the source code of the application,
identifies all its programming constructs and uploads their signatures
as well as an application identifier to the central engine.

Note that the above-described components run at different points in
time. Source Code Analyzer and Runtime Tracer are expected to run
continuously during different phases of the application development
lifecycle and their results are kept even after the release of the
application to its customers. Once the Patch Analyzer is triggered,
typically after the release of a patch for a vulnerable library,
the assessment result is immediately available
thanks to the comparison of the newly collected change-lists with the
previously collected traces.

\section{Proof-of-concept} 
\label{sec:eval}
This section describes our current implementation of the approach and the
architecture presented in Section~\ref{subsec:concept}.
We illustrate its application to CVE-2014-0050 as case study.

\subsection*{Implementation}
\label{subsec:impl}
The current prototype supports the assessment of vulnerabilities of
Java components and is implemented by using technologies that can be
seamlessy integrated in typical Java development and build environments
based on Apache Maven.

The Assessment Engine is realized by means of a SAP Hana database
where the change-list, trace-list and programming constructs are
stored and manipulated. The results are accessible via a web
frontend (see Figure~\ref{fig:list}).

The Patch Analyzer is implemented as a Java stand-alone application. It
interacts with version control systems (VCS) (Git and Subversion in our
current implementation) by means of
the JGit\footnote{\url{http://eclipse.org/jgit/}}
and SVNKit\footnote{\url{http://svnkit.com/}}
Java libraries. Such
libraries are used to
retrieve the code for the patch and vulnerable revisions (i.e., the
revision preceeding the patched one). The ANTLR\footnote{\url{http://www.antlr.org/}}
library
is used for building the parse tree of the Java files to be
compared in order to obtain the change-list.

The Runtime Tracer requires the injection of code into each
programming construct of the application and its libraries. The
instrumentation (both static and dynamic) is realized using
Javassist\footnote{\url{http://www.csg.ci.i.u-tokyo.ac.jp/~chiba/javassist/}}. 
Dynamic instrumentation was found suitable
for unit tests, executed by individual developers and during continuous integration.
Static instrumentation is instead suited for integration
and end-user acceptance tests performed on dedicated systems. In
particular, if the application is deployed in application containers
such as Apache Tomcat, static instrumentations allows to avoid the
performance impact on startup time, which is caused by the significant number of classes loaded before the container and its application become available. Moreover, static instrumentation is also useful in case Java Runtime Environment (JRE) options cannot be accessed or changed to enable dynamic instrumentation, e.g., when using Platform as a Service (PaaS) offerings.
In either cases, whenever the application is executed, the
instrumentation code added is responsible for collecting and uploading
traces to the central engine.  Note that the collection and upload is
only done upon the first invocation of the respective programming
construct which significantly reduces the performance overhead. The
limited impact on application performance has as goal to enable the
trace collection during everyday testing activities.

The Source Code Analyzer is realized by means of a Maven plugin. As
for the Patch Analyzer it uses the ANTLR library to parse the Java
classes. The result is the collection of the signature of every
programming construct belonging to the application itself. The Maven identifier (composed of group id, artifact id, and version) is used as identifier of the application to
set the context for the analysis, i.e., it represents the application
$a$ to define the sets $S_a$ and $T_{a}$ of programming constructs
and trace-list (see Figure~\ref{fig:concept}).
\vspace*{0.6\baselineskip}

\subsection*{Case-study: CVE-2014-0050.} \label{sec:cve}
The National Vulnerability Database (NVD) is a comprehensive, publicly
available database for known vulnerabilities that are disclosed in a
responsible manner, i.e., for which a patch has been made available at
the time of the vulnerability publication. It also includes the
information regarding affected products.
In particular, the Common Vulnerabilities and Exposures (CVE) and
Common Platform Enumeration (CPE) standards are used to identify the
vulnerability and the affected products, respectively.

CVE-2014-0050~\footnote{\url{http://web.nvd.nist.gov/view/vuln/detail?vulnId=CVE-2014-0050}}
describes a vulnerability in Apache FileUpload~\footnote{\url{http://commons.apache.org/proper/commons-fileupload/}} as follows:
\emph{``MultipartStream.java in Apache Commons FileUpload before
  1.3.1, as used in Apache Tomcat, JBoss Web, and other products,
  allows remote attackers to cause a denial of service (infinite loop
  and CPU consumption) via a crafted Content-Type header that bypasses
  a loop's intended exit conditions.''}

Upon disclosure of the vulnerability, any developer using Apache
FileUpload needs to judge whether her application is affected.
Current practices require her to rely on the textual description
for taking a decision. However, it is not straighforward to assess
whether the Java class MultipartStream (referenced to in the
description) is used in the scope of an application.  In fact it may
be used either directly (i.e., instanciated in
the source code of the application) or indirectly within other classes of
the libraries which are directly used.

\begin{figure}[t!]
\centering
\includegraphics[width=.5\textwidth]{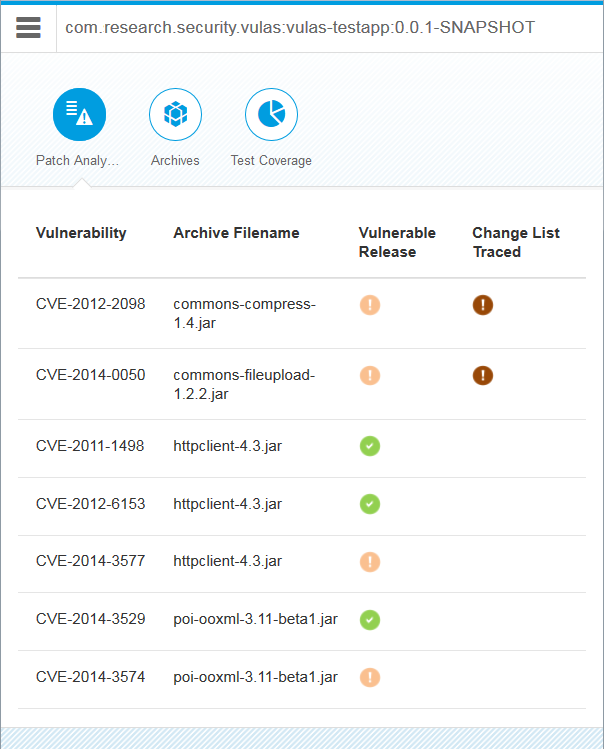}
\caption{Analysis overview for the sample application}
\label{fig:overview}
\end{figure}

The application used in our case study is a sample Web application, com.research.vulas:vulas-testapp:0.0.1-SNAPSHOT, which performs various operations on compressed archives.
Figure~\ref{fig:overview} shows assessment results for several
vulnerabilities after the execution of both JUnit and integration
tests of the Web application as well as the computation of
change-lists. The first two vulnerabilities, including CVE-2014-0050,
are marked as relevant because constructs part of the change-list
(i.e., subject to the security patch) have been executed in its
vulnerable version.  CVE-2011-1498, CVE-2012-6153, and CVE-2014-3529
are marked as irrelevant since non-vulnerable releases of the
respective libraries are used. For CVE-2014-3577 and CVE-2014-3574 the
assessment result shows that the vulnerable release is used but no
traces for the change-list were observed.

\begin{figure}[t!]
\centering
\includegraphics[width=.5\textwidth]{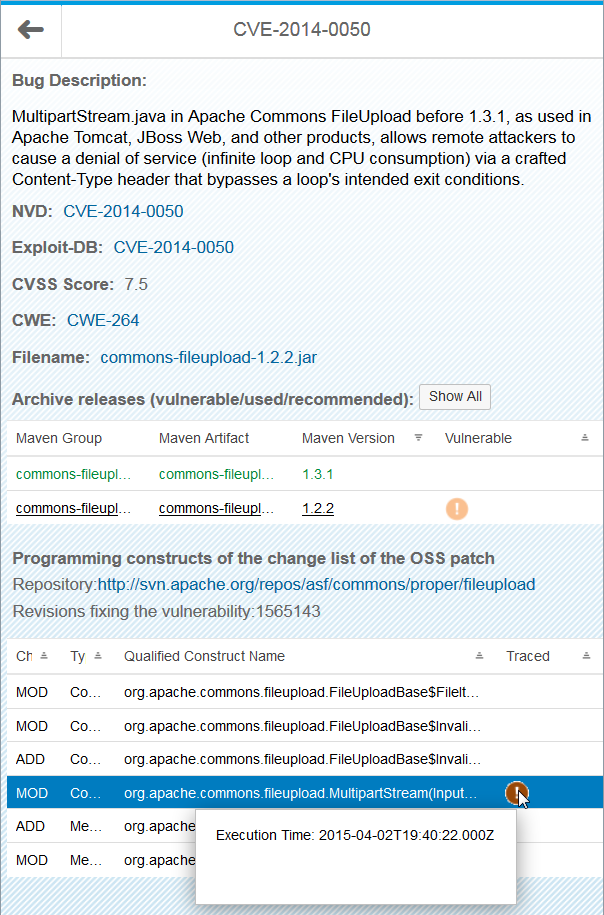}
\caption{Analysis details for CVE-2014-0050}
\label{fig:list}
\end{figure}

More in detail, the table at the bottom of Figure~\ref{fig:list} shows the change-list
 $C_{ij}$ where $i$ is Apache FileUpload and $j$ the vulnerability
CVE-2014-0050. In this case, the intersection $C_{ij}\cap T_a$
is not empty, but contains the constructor of the Java class MultipartStream. The exclamation mark in the ``Traced'' column highlights that 
its execution was observed --at the time shown in the tooltip-- during application tests.

The Patch Analyzer computed the change-list using the URL of the
VCS of Apache
FileUpload 
and the revision number of the patch,
which is provided by NVD in one of the references of CVE-2014-0050.



%
%

As the used programming construct in Figure~\ref{fig:list} was modified as part
of the patch (marked as MOD in the figure), it is present both in the vulnerable
and patched version of the library.
Moreover, the experiment was done for a rather old vulnerability, hence, the
traces are more recent than the security patch itself. As a result, it is
necessary to identify the version of the library in use, and compare it with the
ones affected by the vulnerability.

This is preferably done by means of Maven identifiers and, if this is
not possible, CPEs. In particular, we search the Maven Central
repository (i.e., the default repository for dependency management
with Maven) for the SHA-1 of the archive from which a class was loaded
at runtime and, if a match is found, we obtain the version of the used
library as well as the information about all existing versions.

The products affected by a vulnerability are identified by interacting with the VCS of the respective library. For that purpose, we analyze so-called tags, which are a common means for marking all those repository elements that constitute a given relrease --at least in case of the widely-used versioning control systems Apache Subversion and CVS. In more detail, we identify all tags applied prior to the security patch in question, and parse the Maven project files that existed at that time.

In case of VCS other than Apache Subversion and CVS, we resort to the NVD for establishing the affected versions of a given vulnerability. The NVD uses CPE names, mainly composed of the vendor, product and version information, to identify all affected products. In order to establish if the used library is affected by the CVE we check if its version is among those listed for the affected CPEs. In general, the use of CPE names is considered as a fallback only, since the matching of CPE vendor and product names to Maven identifiers is ambiguous (cf. Section ~\ref{sec:quality}).

Information about library releases is displayed in the upper table of Figure~\ref{fig:list}. It shows that the used library, i.e., apache
commons\_fileupload 1.2.2, is indeed affected by the vulnerability: there existed a corresponding tag that has been applied prior to the commit of the security fix. It also shows the latest,
non-vulnerable release, i.e., apache commons\_fileupload 1.3.1. By
updating to the latest release, the risk of being vulnerable can be
addressed.






As the web application of our case study runs within the
Apache Tomcat application container, we opted for the static
instrumentation in order to avoid the impact of dynamic
instrumentation on the container's initial startup time.
The trace of the patched programming construct
was collected by using integration tests on the interface for
uploading files of the instrumented application. Intuitively, we
perceive that unit and integration tests are complementary means for
collecting traces. In particular, the focus of unit tests on the
business logic of fine-granular components does not cover components
involved in the application's main I/O channels,
many of which rely on OSS libraries, e.g., Apache FileUpload,
HttpClient or Struts.

For CVE-2014-0050 an exploit exists as a Ruby
script in
the Exploit-DB, i.e., an archive of exploits for known
vulnerabilities (\url{http://www.exploit-db.com/exploits/31615/}). By manually running it, we observed that assumption
{\bf (A1)} of Section~\ref{sec:concept} holds in our case study, i.e.,
the vulnerability is exploitable in the given application context even
though only one programming construct belonging to the change-list has
been executed.

\begin{figure}[t!]
\centering
\includegraphics[width=.5\textwidth]{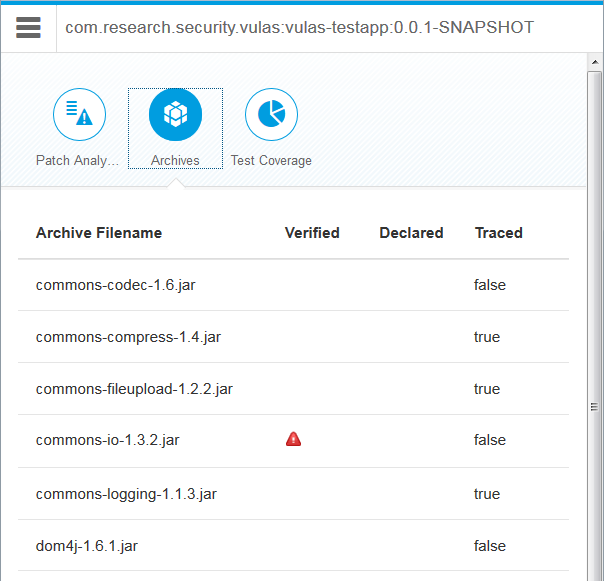}
\caption{Information about archives declared and/or traced}
\label{fig:archives}
\end{figure}

Other than assessing vulnerabilities, the prototype offers two other views:

The first view shows all archives used by the application under analysis (cf. Figure~\ref{fig:archives}), either because they have been declared using Maven or because they have been observed during application tests (i.e., classes where loaded from those archives). Archives whose SHA-1 is not known to the Maven Central are highlighted (commons-io-1.3.2.jar in the example), and so are archives that have not been declared but whose execution was observed during application tests. The former may indicate the use of a tampered archive, the latter bad development practice.

\begin{figure}[t!]
\centering
\includegraphics[width=.5\textwidth]{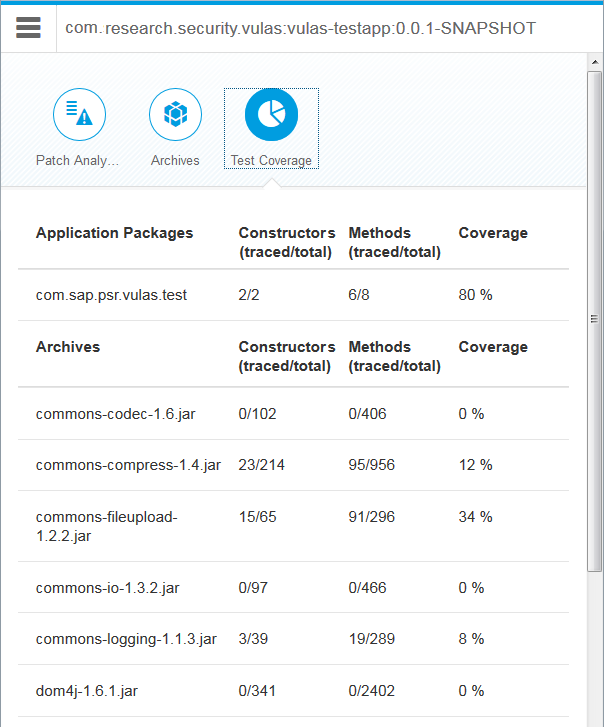}
\caption{Function coverage for application constructs and archives}
\label{fig:coverage}
\end{figure}

The second view displays the function coverage of application constructs as
described in Section~\ref{sec:concept}, aggregated on the level of Java packages (cf. Figure~\ref{fig:coverage}). Moreover, it shows the function coverage for archives used by the application, aggregated on archive level.

\section{Data integration problems}
\label{sec:quality}

Our approach requires the integration of information stemming from
different sources, e.g., vulnerability databases, VCS for managing the
code base of OSS, and public OSS repositories (e.g., Maven
Central\footnote{http://search.maven.org} in case of our Java
prototype).
During our experiments, we found that the integration is hindered by
several problems, each one requiring ad-hoc, technology-specific
solutions. Such problems hamper---in general---the automation of OSS
vulnerability management.

\subsection*{Non-uniform reporting of products affected by a vunerability.} 
\noindent The NVD uses the CPE standard for enumerating components
affected by a vulnerability. In our experiments, we observed a
non-uniform practice of assigning CPEs to vulnerabilities in OSS
libraries. In some cases only the CPE of the respective library is
mentioned. As an example, the affected components for
CVE-2012-2098\footnote{Algorithmic complexity vulnerability in the
  sorting algorithms in bzip2 compressing stream
  (BZip2CompressorOutputStream) in Apache Commons Compress before
  1.4.1 allows remote attackers to cause a denial of service (CPU
  consumption) via a file with many repeating inputs.} are versions of
Apache Commons Compress before 1.4.1.
\begin{verbatim}
cpe:/a:apache:commons-compress
\end{verbatim}
In other cases also the CPEs of applications making use of the library
are listed. This is the case for CVE-2014-0050, whose affected
components include not only Apache FileUpload,
\begin{verbatim}
cpe:/a:apache:commons_fileupload
\end{verbatim}
but also several versions of Apache Tomcat,
\begin{verbatim}
cpe:/a:apache:tomcat
\end{verbatim}
which is just one out of many applications using this library. As
mentioned in the textual description of the CVE
(cf. Section~\ref{sec:cve}), JBoss Web, and other products are also
affected, though, not listed as CPEs. 

The above CPEs are written in the URI binding form. We made use of a
subset of the supported fields, i.e., type \verb|a| to denote an
application, vendor \verb|apache|, product \verb|commons-compress|,
\verb|commons_fileupload|, and \verb|tomcat|. It is important to
notice that CPEs do not provide an immediate means to identify
libraries.

\subsection*{Vulnerability and dependency management make use of different naming schemes and nomenclatures.} 
\noindent There exist many language-dependent technologies and
nomenclatures for identifying libraries and managaging dependencies at
development and build time (e.g., Maven for Java), none of which maps
straight-forwardly to CPE. As an example consider the Apache HTTP
Client library. Maven identifiers are of the form
\begin{verbatim}
GroupId=org.apache.httpcomponents
ArtifactId=httpclient
\end{verbatim}
for each release as of 4.0. CVE-2012-6153 affects Apache Commons
HttpClient before 4.2.3 and the CPEs listed as affected products are
of the form
\begin{verbatim}
cpe:/a:apache:commons-httpclient
\end{verbatim}
While a human can easily recognize the mapping, this is not the case
for an automated solution. The problem is made even worse by the fact
that Maven identifiers and CPEs may change for newer releases. As a
matter of fact the syntax \verb|commons-httpclient| that is used in
the above CPE was also used as Maven group and artifact id for older
releases (before 4.0).

Moreover, there exists no unambigous, language-independent way for
uniquely identifying libraries once they are bundled and installed as
part of an application. Our prototype computes the SHA-1 of Java
archives and performs a lookup in Maven central: an ad-hoc solution
that fails if the bundled library has been compiled with a different
compiler than the library available in the public repository.

\subsection*{Vulnerabilities and VCS information of the respective security patch are not linked in a systematic and machine-readable fashion.} 
\noindent The change-list computation requires as input the URLs of
VCS and commit numbers. The Patch Analyzer of our prototype uses two
strategies for discovering them: \emph{(i)} pattern matching for
identifying VCS information in CVE references and \emph{(ii)} search
for CVE identifiers in VCS commit logs. While successful in case of
CVE-2014-0050 and many other vulnerabilities, they still represent
ad-hoc solutions depending on the discipline of developers and the
quality of CVE entries. In particular, both strategies are successfull
for CVE-2014-0050: \emph{(i)} the VCS
\begin{verbatim}
http://svn.apache.org/r1565143
\end{verbatim}
 is listed among the CVE references;
 \emph{(ii)} the commit log for revision \verb|1565143| contains the CVE identifier:
\begin{verbatim}
Fix CVE-2014-0050. Specially crafted input 
can trigger a DoS if ... This prevents the
DoS.
\end{verbatim}

However in other cases different practices are used. As an example
CVE-2012-2098 affecting Apache Commons Compress does not reference the
VCS revision(s) fixing the vulnerability, nor the VCS commit log
systematically references the CVE.  In this example the revisions
fixing the vulnerability are listed in a webpage of the Commons
Compress project dedicated to security reports
\footnote{http://commons.apache.org/proper/commons-compress/security.html}.

\section{Related work}
\label{sec:relwork}

With the increased adoption of open-source components in commercial products,
the attention to the potential risks stemming from this practice has increased
correspondingly. More specifically, the problem of evaluating the impact of
vulnerabilities of open-source (and more generally, of third-party) components
has attracted significant attention among
researchers~\cite{schryen2011open,arora2006empirical} and
practitioners~\cite{contrast-security,owasptop10}.
Several approaches tackle the problem of vulnerability impact
assessment by examining the system statically, in order to determine a
measure of risk, as in~\cite{brenneman2012improving} where the authors
consider the relation between the entry points of the subject system (the
potential attack surface) and the attack target (the vulnerable code).
Younis et. al~\cite{younis2014using}, elaborating on that idea, proposed a
similar approach that also measures the ratio between damage potential and
attack effort in order to estimate how motivated an attacker needs to be
when targeting a particular point of the attack surface.


Our approach is complementary to these, in that our goal is to observe real
executions of a system (as opposed to analyzing its structure and call graph)
in order to detect whether actual executions were observed that touched a part
of the code that is known to be vulnerable. Judging how likely is the
exploitation of vulnerabilities that were not covered by concrete execution is
not in the scope of this work (although we do plan to include that aspect in our
future research).

Several tools have been proposed to help detect the use of vulnerable libraries,
such as the \textsf{OWASP Dependency-Check}\footnote{\url{https://www.owasp.org/index.php/OWASP_Dependency_Check}}
or the \textsf{Victims Project}\footnote{\url{https://victi.ms/}, \url{https://github.com/victims}}. Both support the
check of whether a project depends on libraries for which there are any known,
publicly disclosed, vulnerabilities. Similarly to ours, these tools are
realized as Maven plugins to minimize the barrier to adoption. They differ from
our approach because their goal is to identify whether a vulnerable library is
\emph{included} in a project, whereas we concentrate on detecting whether the \emph{vulnerable
portion} of the library can be actually \emph{executed} as part of the container
project, a question that is particularly relevant for released applications.

\section{Future directions}
\label{sec:futurework}

Up to now, we have used the current implementation of our approach to a limited
set of sample projects. The evaluation we could make was only preliminary, but
the feedback we received from the the early adopters  of our tool (develoment
units internal to our company) indicates clearly that the problem we are
tackling is perceived as timely and extremely relevant in practice.
That feedback also highlights the importance of several outstanding problems
which demand further investigation. In this section we summarize the future
directions of our research, which will be the topic of future works.

\subsection*{Accuracy of the analysis}
One inherent limitation of our approach, as most existing approaches to
vulnerability analysis, is that it is neither sound nor complete.
In particular, the reliability of our assessment is heavily dependent on the
coverage achieved through executions (e.g., obtained by testing) of the subject
system and its libraries. This has two important consequences: one is related to
the nature of the judgments that one can draw based on testing; the other is
related to the problems that can arise when a test suite constructed for
functional testing is used as the basis of a security assessment.

Obviously, when execution coverage is poor,
it may happen that a vulnerability is not deemed relevant just because no
observed path reached the vulnerable code. This says nothing about
whether such path would be feasible in practice.
Furthermore, even when relying on a well-written (functional) test suite that
achieves high-coverage, there might still be corner cases -- not considered in
functional tests -- that are potentially relevant from a security
standpoint and that an attacker could exploit.

The first point can be addressed by using test suites that achieve good coverage
and that therefore reduce the chances that obvious problematic execution paths
go unnoticed.

Regarding the second point (which can also benefit from testing if
the functional test suite is augmented with explicit tests for corner cases and
negative tests), we feel it would be tackled more effectively by combining our
current test-based approach with static analysis.
By analysing the source code of  the target program and its libraries, we could
determine with some approximation (e.g., constructing the combined call graph),
whether it is at all \emph{possible} to reach vulnerable code from the
application code.

In certain particular cases, this method could provide
very strong evidence that the vulnerable code is \emph{unreachable}, and as such
it would complement nicely our test-based method with provides very strong
evidence (a proof, indeed) in the complementary case, that is when vulnerable
code is indeed \emph{reachable}.
This technique would still need to cope with some degree of approximation.
A study of the interplay of test-based
analysis and static analysis will also require to investigate which types of flow analysis are best
suited to provide a good balance between reliability of the results and
performance,  especially when taking into account the the complexity of large real-world applications.



\subsection*{Scalability to large projects}
While we do not have conclusive evidence nor quantitative figures to offer at this time,
the performance observed in our preliminary tests is promising. 
We believe that the performance penalty that our tool imposes on the build process
would acceptable in most practical cases. As a future work, we plan to conduct a systematic
study of the performance of our tool, by using it in large commercial
applications with complex build structure and hundreds of dependencies.

\subsection*{Experiments to validate the assumptions underlying our approach}
The basic assumption on which this work is based (see Sec.~\ref{sec:intro})
seems sensible based on what we observed in a limited set of sample projects
both in the open-source and in industrial projects. However, its
rigorous validation is a prerequisite for drawing more realiable conclusions
about the quality of our approach. This validation will require examining a
larger number of projects and compare the results of our analysis with the
actual exploitability of vulnerable libraries in the context of those projects.

\subsection*{Tackling the data integration problems}

Our approach heavily relies on data coming from an heterogeneous set of sources,
which include vulnerability databases (such as the NVD) and source code repositories.

Based on our experience, we believe that,
despite the growing attention that both researchers and practitioners dedicate
to the topic of automated vulnerability management, the gap to be filled in is still
quite large. A key problem that  approaches like ours have to face is how to reliably relate
CVE entries with the affected software products and the corresponding source code
repository, down to the level of accurately matching vulnerability reports with the
code changes that provide a fix for them.

This information is currently unavailable, and obtaining it proved to be extremely difficult.

We are currently adoption ad-hoc solutions to these problems. For example, we are manually constructing
a curated list of widely used open-source projects and their respective code repositories.
This approach has the obvious drawback of requiring manual effort both to build and
maintain the list; furthermore the coverage is limited to a large but non-exhaustive
set of projects.  Similarly, we are using ad-hoc mechanisms to determine the correspondence
between release numbers (as mentioned in CVEs) to commit identifiers in source code repositories.

As a future work, we will investigate ways to improve these solutions a we will study
possible alternative methods.

\subsection*{Integrating continuous vulnerability assessment in continuous
integration systems}

Our approach, when considered as part of the overall software development lifecycle, has
a very natural application in continuous build and integration
systems. When included in such systems, our tool can collect traces on a regular basis and therefore
can offer timely notifications when one or more of the used libraries are found
to be affected by a vulnerability report.

At the time of writing, we are initiating the work to adapt our prototype to run
as part of Jenkins builds. As a future work, we intend to complete this
implementation and to evaluate it when used in large development projects (e.g., with over
a hundred libraries).

\section{Conclusion}
\label{sec:concl}

This paper presented a pragmatic approach to answer one important and
time-critical question: Does a vulnerability in bundled OSS libraries
affect an application? 
Our approach helps to assess whether urgent patching is needed in
response to a vulnerability.  It is generic with regard to programming
languages and types of vulnerabilities, and can be seamlessly
integrated into industry-scale build and integration systems.

This paper presented both the conceptual approach and a concrete implementation
as a tool, whose  functionality was demonstrated using an
illustrative example.

\paragraph{Contact information.} The authors can be contacted via e-mail at
their addresses: \texttt{firstname.lastname AT sap.com}. Comments and feedback
on this paper are very appreciated.


\bibliographystyle{IEEEtran}
\bibliography{biblio}

\end{document}